\documentclass[useAMS,usenatbib,a4]{rspublic}
\usepackage[dvips]{graphicx}

\title[Massive stars in their death-throes]{Massive stars in their death-throes}
\author[John J. Eldridge]{John J. Eldridge \thanks{E-mail: jje@ast.cam.ac.uk} \\
Institute of Astronomy, The Observatories, University of Cambridge, Madingley Road, Cambridge, CB3 0HA\\}

\begin{document}
\maketitle
\label{firstpage}

\begin{abstract}{stars: evolution --  supernovae: general -- stars: Wolf-Rayet -- stars: supergiants}

  The study of the stars that explode as supernovae used to be a
  forensic study, working backwards from the remnants of the
  star. This changed in 1987 when the first progenitor star was
  identified in pre-explosion images. Currently there are 8 detected
  progenitors with another 21 non-detections, for which only a limit
  on the pre-explosion luminosity can be placed. This new avenue of
  supernova research has led to many interesting conclusions, most
  importantly that the progenitors of the most common supernovae, type
  IIP, are red supergiants as theory has long predicted. However no
  progenitors have been detected thus far for the hydrogen-free type
  Ib/c supernovae which, given the expected progenitors, is an
  unlikely result. Also observations have begun to show evidence that
  luminous blue variables, which are among the most massive stars, may
  directly explode as supernovae. These results contradict current
  stellar evolution theory. This suggests that we may need to update
  our understanding.

\end{abstract}

\section{Introduction}

The term supernovae was first introduced by Baade \& Zwicky (1935)
when they separated such events from \textit{common} novae.  Novae
occur frequently in our galaxy and are the result of hydrogen
accreting on to the surface of a white dwarf star that ignites after a
thick enough layer is deposited. The \textit{super}-novae are far
more luminous and rare, only occurring once or twice a century in our
own Galaxy. The last supernovae observed in our Galaxy were the Tycho
and Kepler Supernovae in 1572 and 1604 respectively.

Supernovae can be so luminous that they outshine all the other stars
in a galaxy. After a sharp rise to maximum light over a few days their
luminosity decays and slowly fades over weeks and months. Some have a
light curve plateau and remain at a constant brightness for up to four

months before fading. The energy required to power a typical supernova
display is similar to the amount of energy our Sun will output over
its 10 billion year lifetime. To produce that amount of energy over a
few days the progenitor star must be significantly altered in a
supernova.

\begin{figure}
\begin{center}
\includegraphics[angle=0, width=100mm]{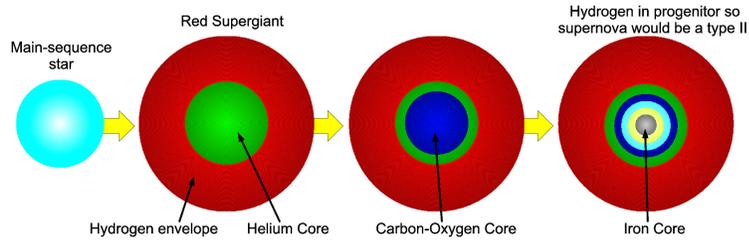}
\caption{A massive star's life cycle. Stars begin their lives on the
  main sequence as compact blue stars. Once the helium core is formed
  the surrounding hydrogen envelope expands and the star becomes a red
  supergiant. Further burning stages occur until an iron core is
  formed. The star will explode as a type-II supernova. }
\label{single}
\end{center}
\end{figure}

The modern understanding of supernovae divides them into two main
types, thermonuclear and core-collapse. Thermonuclear supernovae,
(also know as type Ia supernovae) are the detonation of carbon-oxygen
white dwarfs. We do not discuss them in this review (see Hillebrandt
\& Niemeyer 2000). Core-collapse supernovae account for 72 percent of
all supernovae in a volume limited sample (Smartt et al. 2008). They
are the final events in the lives of stars more massive than
approximately 8 solar masses ($8M_{\odot}$). The basic evolution of a
star is relatively well understood. A star begins on the main
sequence, burning hydrogen to helium. Here it spends the largest
fraction of its lifetime. Once a helium core is formed, hydrogen
continues to burn in a shell around the core and the stellar radius
swells to between 100 and 1000 times that of the Sun. At this stage
the star becomes a red giant or, for the most luminous, a red
supergiant. Massive stars undergo further burning stages. Helium burns
to form carbon and oxygen and then progressively heavier elements
burn, preventing stellar collapse until an iron core is formed. Iron
fusion is an endothermic reaction. With no further energy source to
prevent the core from collapsing a neutron star or a black hole is
formed. Figure \ref{single} shows the evolution of a star in schematic
form.

Creating a neutron star releases a tremendous amount of energy in
neutrinos. A fraction of these neutrinos interact with the stellar
envelope, providing the energy to heat and eject it and thus give rise
to the supernova. The ejected material is rich in heavy elements
synthesized during the star's life and in the supernova. Eventually
this material mixes with the interstellar medium and pollutes it. When
future generations of stars form they will be more metal rich than the
previous generation. Most of the heavy elements in our bodies was formed in
the evolution of a massive star and its explosive end.

The exact mechanism of how the energy is transfered to the envelope is
uncertain. Most current simulations of supernova do not produce
explosions after the core becomes a neutron star. Different additional
mechanisms such as acoustic driving by the proto-neutron star or jets
from a magneto-rotational instability as material accretes on to the
proto-neutron star have been suggested (see Burrows et al. 2007,
Dessart et al. 2008 and references therein).

There are two exceptions to the standard picture of iron core
collapse. Stars around 7 to $8M_{\odot}$ are not massive enough to
progress past carbon burning they therefore form oxygen-neon-magnesium
cores supported by electron degeneracy pressure. If the core mass
reaches the Chandrasekhar mass of $1.4M_{\odot}$ it begins to
collapse. Eventually the central density is high enough that electrons
are captured by magnesium and neon. This removes the electrons
supporting the core and accelerates the collapse to a neutron star and
an electron-capture supernova occurs (Eldridge, Mattila \& Smartt
2007; Poelarends et al. 2008). The other exception is the
pair-production instability when photons in the core form
electron-positron pairs reducing the radiation pressure that was
supporting the star and hence leading to its collapse. The resulting
evolution is complex and the entire star can be disrupted in an
explosion (Heger et al. 2003). However such supernovae are rare,
occurring only in the most metal poor galaxies (Langer et al. 2007).

\begin{figure}
\begin{center}
\includegraphics[angle=0, width=100mm]{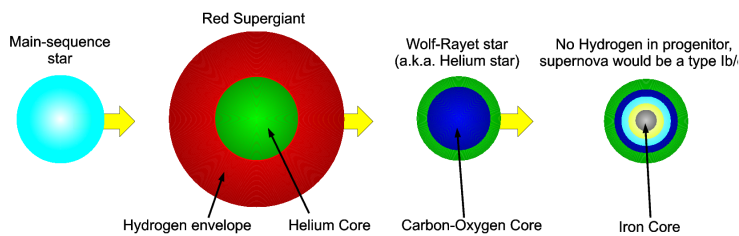}
\caption{The details of the figure are similar to Figure
  \ref{single}. However after the star becomes a red supergiant strong
  stellar winds (or binary interaction) remove the hydrogen envelope
  and the star becomes a Wolf-Rayet star. Once an iron core is formed
  the star explodes as a type-Ib/c supernova.}
\label{single2}
\end{center}
\end{figure}

The appearance of a supernova strongly depends on the structure and
composition of the material ejected. This is determined by how much
mass is lost from the surface of the star during its evolution. Stars
of mass less than 25$M_{\odot}$ have weak stellar winds and do not
lose their hydrogen envelopes before an iron core is formed. Stars
more massive than this have strong stellar winds and all hydrogen is
lost from the surface before the star explodes. These stars are
known as helium stars or Wolf-Rayet stars and Figure \ref{single2}
shows the structure of such a star. The hydrogen envelope can also be
removed if the star is in a binary. The stars in a binary interact if
their radius is similar to the radius of the orbit and mass-transfer
can occur with one star losing mass and the other gaining mass.

Core-collapse supernova progenitors naturally separate into two main
groups, those that contain hydrogen when they explode and those that
do not. Supernovae are also classified by the same aspect. If hydrogen
is detected in the supernova spectrum then it is classified as a
type-II supernova otherwise it is a type-I. These two main types can
be further subdivided based on the photometric and spectroscopic
behaviour of the supernova (Filippenko 1997).

There are two types of type-I core-collapse supernovae:
Type Ib with helium lines but no silicon lines in their
spectra and Type Ic with neither helium nor silicon lines in
their spectra (silicon lines indicate a thermonuclear type Ia supernova).

For type-II supernovae, if the luminosity is constant for a few months
after the supernova appeared, a plateau in the light curve, then it is
a type IIP. These are the most common type of supernovae, making up 59
percent of core-collapse supernovae in a volume limited survey (Smartt
et al. 2008). The luminosity is constant because, as the ejecta expand
in radius, the visible surface where the observed light is emitted from
(the photosphere) moves inwards in mass and therefore remains stationary.

There are three other rarer subtypes of type-II supernovae. If the
light curve decays linearly then the supernova is a type IIL. The
progenitors of these stars are thought to have lost hydrogen
from their envelopes so that there is not enough to produce
the luminosity plateau. If the supernova has narrow hydrogen lines
which indicates slow moving ejecta it is a type IIn. This occurs when
the supernova ejecta encounter a large amount of material surrounding
the progenitor star and are decelerated from a typical ejection
velocity of $10^{5} {\rm km\,s^{-1}}$ to $1000 \, {\rm km\,s^{-1}}$. The
last subtype is the hybrid class, type IIb. These are type-II events
which metamorphose into Ib supernovae. The progenitors of these
supernovae have only the barest trace of hydrogen left at the time of
core collapse.

The deduction of which stars produce which supernova was a series of
educated guesses. It was thought that red supergiants produced
type-IIP supernovae and that the other types were produced depending
on the amount of mass lost before core collapse. The more mass that is
lost from a star, the deeper the layers and the heavier elements that
are exposed at the surface. This leads to the stellar type of the
progenitor changing from a red supergiant to a Wolf-Rayet star and the
supernova type progresses from: IIP $\rightarrow$ IIL $\rightarrow$ IIb
$\rightarrow$ Ib $\rightarrow$ Ic. The amount of stripping depends on
the initial mass of the star with the most massive stars losing the
largest fraction of their initial mass and exposing the deepest
interiors at explosion.

The only way to confirm whether this theory is correct is to study the star
that exploded in a supernova. This is difficult because the star is
destroyed in the supernova. It is possible however that an image may
have been taken before the explosion but supernovae are rare. Also only
a few galaxies are close enough for individual stars to be resolved in
ground-based images. However the Supernovae, 1987A and 1993J, took
place in nearby galaxies and it was possible to find the progenitor
star in pre-explosion images and determine their nature (see Section
2).

The situation dramatically improved with the launch of the Hubble
Space Telescope (HST). Its resolving power and sensitivity were such
that it became possible to resolve individual stars in galaxies out to
60 million light-years rather than out to a few million
light-years. This increased the number of galaxies observed in detail
and for a few supernovae a year pre-explosion images of the progenitor
should exist (rather than a few per century before HST). In 2003 two
groups simultaneously found the first progenitor of a type IIP
supernova (Van Dyk et al. 2003; Smartt et al. 2004). With the concept
tested, many discoveries have followed. The success can be reflected
in that, out of the 135 supernovae that occurred from 1999 to 2007
within range of HST, 29 had HST pre-explosion images of the supernova
site. The results have confirmed some theories and caused problems for
others.

In this article we review the study of supernova progenitors,
beginning with the first two discovered. This is followed by
discussion of the study of progenitors during the HST era, with
highlights of the main results of the observations of type-IIP
progenitors. We also discuss the progenitors for type-Ib/c supernovae
and the growing evidence that type-IIn supernova have progenitors that
challenge the preconceptions of stellar theorists.

\section{The Supernovae 1987A and 1993J}

The first two supernovae with observed progenitors were rare
types. In both cases, the progenitor was the result of binary,
rather than the better-understood single star, evolution.

\begin{figure}
\begin{center}
\includegraphics[angle=0, width=100mm]{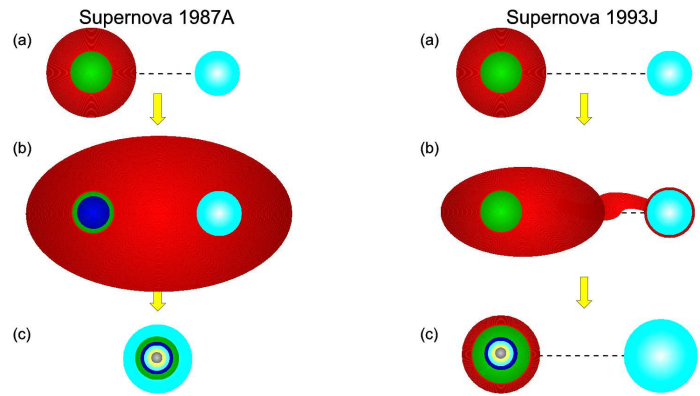}
\caption{Evolution models of the progenitors of Supernovae 1987A and
  1993J. For 1987A (a) the more massive star became a red supergiant,
  (b) the star expanded and engulfed its companion in its hydrogen
  envelope, common envelope evolution occurred, (c) the companion was
  absorbed into the envelope of the primary star and the resulting
  star was a blue supergiant. The core then continued to evolve and
  the star exploded as a blue supergiant in an unusual type-II
  supernova. For 1993J (a) the more massive star became a red
  supergiant first, (b) the envelope grew until surface material was
  attracted by the companions gravity and mass transfer occurred, (c)
  the donating star shrank and mass transfer ended. The companion star
  was then the more massive. The progenitor retained only a small amount of
  hydrogen and so a type-IIb supernova occurred.}
\label{binary}
\end{center}
\end{figure}

\subsection{Supernova 1987A}

Supernova 1987A occurred in the Large Magellanic cloud, a satellite
galaxy of the Milky way. This made it the closest event to be observed
since the invention of the telescope and it was also visible to the
naked eye. It was an unusual supernova, spectroscopically similar to a
type-IIP supernova but with a peculiar light curve. The progenitor was
discovered from photometry and spectroscopy to be a blue supergiant
with a small radius, 45 times the radius of the Sun (Walborn et
al. 1987), whereas theory predicted it should have been a red
supergiant with a radius a few hundred times that of the Sun. We now
understand the supernova was unusual because of the small radius of
the progenitor but why was the progenitor blue? There are several
possible reasons, including low metallicity, rapid rotation or binary
evolution (Podsiadlowski 1992). 

The favoured hypothesis today is that the progenitor was the result of
two stars merging in a binary. Initially both stars were on the main
sequence burning hydrogen to helium. The more massive, a $16M_{\odot}$
star, burnt all core hydrogen to helium first and expanded to become a
red supergiant. Then between this point and core collapse the size of
the primary star became greater than the radius of the orbit and the
whole system entered a common-envelope phase of evolution. The
secondary star, a $3M_{\odot}$ star, was swallowed by the more massive
primary to form a single more massive blue supergiant.

While this model agreed with the progenitor observation, there was no
other evidence supporting the theory of binary evolution. Then, in
1997 a triple-ring system that had surrounded the progenitor became
visible after it was ionized by the supernova's ultraviolet flash. Analysis
showed that these rings were formed, during the common-envelope phase
of evolution, from material that was lost during the merger.  This
was further evidence for the binary scenario and provided a method to
determine that the merger occurred 20,000 years before the supernova
(Morris \& Podsiadlowski 2007).

\subsection{Supernova 1993J}

Supernova 1993J started out as a type-IIb supernova. The progenitor
must have lost most, but not all, hydrogen from its envelope before
core collapse. The event was nearby in the galaxy M81, 12 million
light-years away. Pre-explosion images of the object were consistent
with a red supergiant but there was excess blue flux (Aldering et
al. 1994). The immediate suggestion put forward was that the
progenitor was in a binary system and had a blue companion star. In
2004 the supernova had faded enough for its position to be observed by
HST. The blue companion star was found but the red supergiant had
disappeared confirming that red supergiant star had exploded and a
binary companion was present (Maund et al. 2004).

The binary companion is necessary because, without it, the exploding
star would have retained much more hydrogen and have produced a
type-IIP supernova. Initially the progenitor was $15M_{\odot}$ but
lost $10M_{\odot}$ of material. After it became a red supergiant and
its radius became similar to the orbital separation. Unlike the
extreme interaction of 1987A, material was transferred to the companion
which increased in mass from $14M_{\odot}$ to $22M_{\odot}$, the
remainder being lost from the system (Maund et al. 2004). In the future
this star will also explode but it is difficult to predict how the
accretion will have affected the resulting future evolution.

\section{Supernova 2003gd and other type IIPs}

\begin{figure}
\begin{center}
\includegraphics[angle=0, width=110mm]{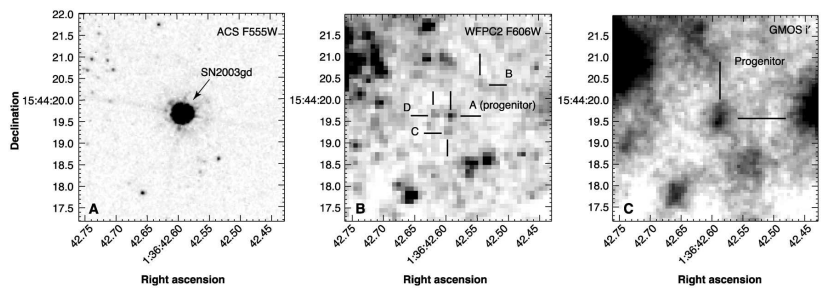}
\caption{The progenitor discovery images for supernova 2003gd (Smartt
  et al. 2004; reprinted with permission from AAAS). Left: the HST post-explosion image showing the
  location of the supernova in relation to nearby stars. Centre:
  the HST pre-explosion image showing the location of the
  progenitor. Right: a ground based Gemini telescope image
  showing the reason why the HST is required to make these
  observations. If this image had been used, the star identified as the
  progenitor would have been a blend of stars A and C in the centre
  panel.}
\label{03gd}
\end{center}
\end{figure}

After HST was launched in 1990 its archive grew and so did the chance
that a supernova would have a pre-explosion image. The first
progenitor discovered by HST was the red supergiant progenitor of
Supernova 2003gd. The star was remarkable in its normality (see
Figure \ref{03gd}, Van Dyk et al. 2003; Smartt et
al. 2004). Subsequently a firm progenitor detection was made for
Supernova 2005cs (Maund et al. 2005; Li et al. 2006). The two
observations confirmed for the first time that the progenitors of the
most common type of supernova were the expected red supergiants, as
shown in Figure \ref{single}. Both were found to have masses of
$8M_{\odot}$, the predicted minimum mass for a supernova to occur
(e.g. Heger et al. 2003; Eldridge \& Tout 2004).

Other supernovae with available pre-explosion images have less
conclusive detections and in many cases no progenitors have been
detected. However it is possible to place an upper limit on how
luminous (and thus massive) the progenitor could have been and yet
remain undetected

For the IIP progenitors enough observations are available (20
detections and non-detections) to make some statements about the
progenitor population, specifically the mass range of stars that gives
rise to these supernovae. A statistical analysis shows that the
minimum mass for a star to explode in a supernova is around
$7.5M_{\odot}$, while the maximum mass to give rise to a type-IIP
supernova is around $16.5M_{\odot}$ (Smartt et al. 2008). Therefore we
know that type-IIP supernovae only come from a relatively small range
of masses, despite being in 59 percent of core-collapse
supernovae.

There is a problem that theory suggests that single stars with masses
between 16.5 to 25M$_{\odot}$ should retain their hydrogen envelopes.
So what do these stars explode as? One answer is that the hydrogen
envelope is not massive enough to produce the plateau in the light
curve resulting in a type-IIL or IIb supernova. An alternative is that
these stars have cores massive enough for a black hole to form at core
collapse and have only a small explosion energy because a large
fraction of ejecta material fall back on to the remnant. In principle
the resulting supernovae may be dim and difficult to observe. Until
progenitors for the other type II supernovae are observed we shall
continue to speculate.

\section{Type-Ib/c supernovae}

A massive star can lose its hydrogen envelope in stellar winds or a
binary interaction. The resulting Wolf-Rayet stars are
the suspected progenitors of type-Ib/c supernovae (see Figure
\ref{single2}). It is difficult to separate out the type-Ib and Ic
progenitors as both stellar models and observed Wolf-Rayet stars tend
to contain helium. In general the more helium-rich Wolf-Rayet stars
produce type-Ib supernovae while the highly stripped helium-poor
Wolf-Rayet stars produce type-Ic supernovae.

There are currently nine type-Ib/c supernovae with pre-explosion
images but their progenitors are undetected (Crockett et
al. \textit{in prep.}). If we assume the observed Wolf-Rayet
population (van der Hucht 2001) are the progenitors of these
supernovae and use the detection limits from the progenitor
observations to run a Monty-Carlo simulation we find the probability
of nine non-detections is less than 0.05. This suggests that
Wolf-Rayet stars are not the only type of progenitor. The most
sensitive pre-explosion observation to date is for Supernova
2002ap. The observation rules out a normal Wolf-Rayet star and favours
a binary system with a \textit{low-mass Wolf-Rayet star} with lower
mass $M\le 5M_{\odot}$, than a typical Wolf-Rayet star, $M \approx
10M_{\odot}$ (Crockett et al. 2007).

Low-mass Wolf-Rayet stars lose their hydrogen envelopes in binary
interactions. These are stars with an initial mass less than
25$M_{\odot}$ which cannot become Wolf-Rayet stars by
themselves. However such stars have never been observed in our
Galaxy. They may be difficult to find because they are less luminous the
normal Wolf-Rayet stars or the binary companion required to strip the
hydrogen may hide the low-mass Wolf-Rayet star. Nothing can be firmly
concluded until a progenitor is observed. Even if the explosion of a
normal Wolf-Rayet star is observed, there are still the non-detections
to explain. Hence, the low-mass Wolf-Rayet stars in the Galaxy must be
found to back up this hypothesis. The only uncertainty is will we be
able to recognize them when we observe them?

\section{Type-IIn supernovae have LBV progenitors?}

A supernova is classified as a type IIn when there are narrow hydrogen
lines in its spectrum. This indicates that the hydrogen is moving
slower than the typical ejecta velocities. These lines occur when the
ejecta are slowed through interaction with dense circumstellar material
around the progenitor star, such as a dense stellar wind. The stars
with the densest winds are luminous-blue variable (LBV) stars. Rather
than having a steady constant wind, these stars are highly variable
and can eject more than a solar mass in a single mass-loss event, a
process that leads to a very dense environment around the
star. Traditionally they have been considered transition objects
between main-sequence stars and Wolf-Rayet stars with initial mass
greater than 60$M_{\odot}$.

The first time LBVs were suggested to be the progenitors of some
supernovae was by Kotak \& Vink (2006). They were able to model
modulations in radio lightcurve of some supernovae by assuming the
pre-supernova mass-loss varied as for an LBV star undergoing S-Doradus
type variations. Then supernova 2005gl increased interest further in
this supernova type and their progenitors. A source consistent with an
LBV was in pre-explosion imaging but it is too distant to be sure it
was a single star and not a cluster of stars (Gal-Yam et al. 2007). In
addition the Supernovae 2006jc, 2006gy and 2005gj had indirect
evidence that their progenitors had properties similar to LBV stars
(Pastorello et al. 2007; Smith et al. 2007; Trundle et al. 2008). This
is an active area of research and the subject is in a state of
flux. LBV stars have traditionally been interpreted as objects that
have yet to begin or complete core helium burning. Suggesting that
they can explode is uncomfortable for most theorists because the stars
still have to burn helium and the other products before
core-collapse. While these burning stages occur the envelope can be
lost and the star becomes a Wolf-Rayet star.

A possible approach is to say that most LBVs are still transition
objects. We therefore need to ask if there are stars that look like
LBV stars but are in fact close to core-collapse. To answer this
question we would need to understand what drives the LBV behavior. We
do not fully understand this behavior although Stothers \& Chin (1996)
find that, in the most massive stars with hydrogen envelopes and
helium cores, the stellar structure separates into two quite separate
structures, an inner core and a separate unstable outer shell. It is
this outer shell that is ejected from the star, leaving the inner core
intact. These structures occur in the most massive stars after the end
of the main sequence. However a similar structure can be found in a
narrow mass range, around $30M_{\odot}$, in more evolved stars near to
core collapse. While typical LBV stars experience the evolution
sequence, Main sequence $\rightarrow$ LBV $\rightarrow$ Wolf-Rayet
$\rightarrow$ Supernova, these more centrally evolved LBVs experience
the evolution of Main sequence $\rightarrow$ Red supergiant
$\rightarrow$ LBV $\rightarrow$ Supernova (Eldridge, \textit{in
  prep.}). During the red-supergiant phase their cores evolve close to
core collapse as their luminosities grow. They become LBV stars at the
same time as their cores are close to collapse. These stars would
differ from the traditional LBVs in their evolutionary status but
would appear as LBV stars when observed and be indistinguishable. The
luminous and variable red supergiant HV 11423 (Massey et al. 2007)
could be a red supergiant evolving towards an LBV phase of evolution .

Finally a recent type-IIn Supernova 2008S must be mentioned. While no
progenitor has been observed for this object, a dusty cocoon
surrounding the progenitor star was detected. The inferred luminosity
indicates that the progenitor was a star with a mass 4 to $7M_{\odot}$
rather than a LBV star (Prieto et al. 2008). However there is some
debate as to whether the event was a supernova because several
expected observational features have not been observed.

The true nature of LBV stars and the progenitors of type-IIn
supernovae make a complex problem, investigation of which is currently
observationally led. More work by theoreticians is required in
addition to further detailed observational study.

\section{Discussion \& Conclusions}

The direct study of pre-supernova imaging in the archives has led to
some quite remarkable and usually understated successes in the
determination that the progenitors of type IIP supernovae are red
supergiants. Despite this, it is still a subject in its infancy. The
remaining supernova types are still lacking firm constraints. The
importance of direct study is highlighted by the current confusion
of type-IIn progenitors.

By detecting the progenitors of supernovae, we can make fundamental
tests of stellar evolution that were not possible before we were able
to observe their progenitors. Each time we see something new it
provokes us to reach a new understanding. Our understanding of the
lives of stars and the nature of supernovae is evolving dramatically
because of this exciting avenue of science.

\begin{acknowledgements}
  JJE is funded by the IoA Theory Rolling Grant from the STFC. He also
  thanks Julie Wang, Richard Stancliffe, Elizabeth Stanway,
  Christopher Tout and the two anonymous referees for the useful and
  helpful guidance and comments. He also thanks Stephen Smartt and Mark Crockett on continuing collaboration on supernova progenitors.
\end{acknowledgements}

\label{lastpage}

\end{document}